\documentclass[aps,pre,%
preprint,%
amssymb,amsmath,%
nopreprintnumbers,%
showpacs,%
showkeys,%
fleqn,%
eqsecnum,%
byrevtex]{revtex4}
\usepackage{calc}
\usepackage[final]{graphicx}
\begin{document}
\title{The New Screening Characteristics of Strongly Non-ideal and Dusty Plasmas.
Part 3: Properties and Applications}
\author{A. A. Mihajlov$^{1}$}
\email{mihajlov@phy.bg.ac.rs}
\author{Y. Vitel$^{2}$}
\email{yv@ccr.jussieu.fr}
\author{Lj. M. Ignjatovi\'c$^{1}$}
\email{ljuba@phy.bg.ac.rs}

\affiliation{$^{1}$Institute of Physics, P.O. Box 57, 11080 Zemun,
Belgrade, Serbia} \affiliation{$^{2}$Laboratoire des Plasmas Denses,
Universite P. et M. Curie, 3 rue Galilee, Paris, 94200 Ivry sur
Seine, France}
\begin{abstract}
The physical sense and the properties of a group screening
parameters which are determined in the previous papers (Part~1 and
Part~2) for single- and two-component systems is discussed in this
paper. On the base of data from the mentioned papers are determined
two new characteristic lengths which complete a new hierarchy system
of screening lengths in plasma. It was shown that the methods
developed in Part~1 and Part~2 generates results which are
applicable to the strongly non-ideal gaseous and dusty plasmas, and
manifest a very good agreement with the existing experimental data.
\end{abstract}
 \maketitle
\section{Introduction}
\label{sec:intro}

In the previous papers \cite{mih08,mih08a}, here Part~1 and Part~2,
are developed the methods for describing of the electrostatic
screening in single- and two-component systems (gas of electrons on
a positive charged background, some electron-ion and dusty plasmas
etc.). These methods generate a group of the new screening
parameters which characterize the considered systems. In this paper
the physical sense of these parameters is discussed, and the
reinterpretation of some known screening parameters (Landau radius,
non-ideality parameters etc.) is given. On the base of results from
Part~1 and Part~2 in this paper are obtained two new screening
parameters which provide the possibility to form a new hierarchy
system of the characteristic screening lengths and to compare the
obtained results with the existing experimental data.

The material presented in this paper is distributed in four Sections
and one Appendix. In Section~\ref{sec:rem} are given the some
quantities and relations from Part~1 and Part~2 which should make
easier reading of this paper; in Sections~\ref{sec:nip1} and
\ref{sec:nip2} are considered "small" characteristic lengths and
connected with them non-ideality parameters for single- and
two-component systems. Besides, in the same Sections are introduced
"medium" and "large" characteristic lengths for the considered
systems. The obtained results are compared with the existing
experimental data in Section~\ref{sec:nip2}. The conclusions of this
paper are given in Section~\ref{sec:conc}. Finally, one important
example of the application of obtained results, related to the
systems with more than two components, is considered in
Appendix~\ref{sec:scon}.

\section{Remarks: the main quantities and relations}
\label{sec:rem}

\subsection{Single-component systems}
\label{sec:rem1}

The density, temperature and the charge of free particles in the
initial (homogeneous) system: $N$, $T$ and $Ze$, where $Z=\pm 1, \pm
2 ...$, and $e$ - the modulus of the electron charge;

The screening constant:
\begin{equation}
\label{eq181}  \kappa \equiv \frac{1}{r_\kappa}=\left[ \frac{4\pi
(Ze)^2}{\partial \mu/\partial N} \right]^{1 \over 2},
\end{equation}
where $\partial \mu/\partial N \equiv \partial \mu(N,T)/\partial N$,
$\mu(N,T)$ is the chemical potential, and $r_\kappa$- the
corresponding characteristic length;

The charge density in the accessory (non-homogeneous) system with
the probe particle:
\begin{equation}
\label{eq21}  \rho (r) = \rho_{b} + Ze \cdot n(r),
\end{equation}
where $\rho_{b}= - Ze \cdot N$ is the background charge density, and
$n(r)$ - the free particle density at the distance $r$ from the
origin of the chosen reference frame;

In the region $r \ge r_{0}$ we have that
\begin{equation}
\label{eq21a}  \rho (r)=-Ze N\cdot r_0 \cdot \exp (\kappa r_0 )
\cdot \frac{\exp ( - \kappa r)}{r}.
\end{equation}

The radius $r_0$ and the parameters $\gamma _{s}(x)$ and
$\gamma_{\kappa }(x)$:
\begin{equation}
\label{eq261}  r_0 = r_s \cdot \gamma _{s}(x) , \qquad r_0 =
r_\kappa \gamma _{\kappa }(x),
\end{equation}
\begin{equation}
\label{eq251}  \gamma _{s}(x) = [(1 + x^3)^{\textstyle{1 \over 3}} -
1]/x, \qquad \gamma _{\kappa }(x) = (1 + x^3)^{1 \over 3} - 1,
\end{equation}
\begin{equation}
\label{eq271}  x = \kappa r_s = \frac{r_s}{r_{\kappa }}, \qquad
r_{s}=\left(\frac{3}{4 \pi N} \right)^{\frac{1}{3}}
\end{equation}
where $r_{s}$ is Wigner-Seitz's radius (the radius of the sphere
with the volume $1/N$);

The condition of electro-neutrality of the accessory system:
\begin{equation}
\label{eq41}  Ze + \int\limits_0^\infty {\rho (r)} \cdot 4\pi r^2dr
= 0,
\end{equation}
where $Ze$ is the charge of the fixed probe particle.

\subsection{Two-component systems}
\label{sec:rem2}

The density, temperature and the charge of ions in the initial
(homogeneous) system: $N_{i}$, $T_{i}$ and $Z_{i}e$, where $Z_{i}=
1, 2 ...$;

The density, temperature and the charge of electrons in the same
system: $N_{e}$, $T_{e}$ and $-e$ or $Z_{e}e$, where $Z_{e}= -1$;

The ion and electron screening constants and the characteristic
lengths:
\begin{equation}
\label{eq292}  \kappa _{i,e} \equiv \frac{1}{\kappa _{i,e}} =
\kappa_{0;i,e}
 \cdot \left( {1 - \alpha } \right)^{\textstyle{1
\over 2}}, \qquad \kappa_{0;i,e} = \left[ \frac{4\pi (Z_{i,e} e)^2}
{\partial \mu_{i,e}/\partial N_{i,e} } \right]^{1 \over 2},
\end{equation}
where $\partial \mu_{i,e}/\partial N_{i,e} \equiv \partial
\mu_{i,e}(N_{i,e},T_{i,e})/\partial N_{i,e}$, $\mu_{i}(N_{i},T_{i})$
and $\mu_{e}(N_{e},T_{e})$ are the ion and electron chemical
potentials, $\alpha$ is the electron-ion correlation coefficient
given by expression
\begin{equation}
\label{eqalf} \alpha = 1 - \frac{\textstyle{2 \over 3}x_{s}^3
}{\left( {1 + x_{s} } \right)\exp \left( { - x_{s} } \right) -
\left( {1 - x_{s} } \right)\exp \left( {x_{s} } \right)},
\end{equation}
and $x_{s}=\kappa_{0;e}r_{s;i}$;

The charge densities in the accessory (non-homogeneous) systems with
the probe particles:
\begin{equation}
\label{eq22}  \rho ^{(i,e)}(r) = Z_i e \cdot n_i^{(i,e)} (r) - e
\cdot n_e^{(i,e)} (r),
\end{equation}
where $n_i^{(i,e)} (r)$ and $n_e^{(i,e)} (r)$ are the ion and
electron densities at the distance $r$ from the origin of the chosen
reference frame; upper indexes $(i,e)$ pointed the considered case:
$(i)$ - the probe particle charge is equal to $Z_{i}e$, $(e)$ - the
probe particle charge is equal to $-e$;

In the region $r \ge r_{s;i,e}$ we have that
\begin{equation}
\label{eq21b}  \rho^{(i,e)} (r)= Z_{e,i}e (1-\alpha)\cdot N_{i,e}
r_{0;i,e} \cdot \exp (\kappa_{i,e} r_{0;i,e} ) \cdot \frac{\exp ( -
\kappa_{i,e} r)}{r},
\end{equation}
where $Z_{e}=-1$.

The radii $r_{0;i,e}$ in the cases $(i)$ and $(e)$ and the
parameters $\gamma _{s}(x_{i,e})$ and $\gamma_{\kappa}(x_{i,e})$:
\begin{equation}
\label{eq702}  r_{0;i,e} = \gamma_{s}(x_{i,e}) \cdot r_{s;i,e},
\qquad r_{0;i,e} = \gamma_{\kappa}(x_{i,e}) \cdot r_{\kappa;i,e}
\end{equation}
\begin{equation}
\label{eq712}  x_{i,e} = \kappa_{i,e}
r_{s;i,e}=\frac{r_{s;i,e}}{r_{\kappa;i,e}}, \qquad
r_{s;i,e}=\left(\frac{3}{4 \pi N_{i,e}} \right)^{\frac{1}{3}}
\end{equation}
where $r_{s;i,e}$ are Wigner-Seitz's radii (the radii of the spheres
with the volumes $1/N_{i}$ and $1/N_{e}$);

The electro-neutrality conditions of electro-neutrality of the
accessory systems:
\begin{equation}
\label{eq42}  Z_{i,e} e + \int\limits_0^\infty {\rho ^{(i,e)}(r)}
\cdot 4\pi r^2dr = 0,
\end{equation}
where $Z_{i,e} e$ is the charge of the corresponding probe particle.

\section{The screening parameters of the single-component systems}
\label{sec:nip1}

\subsection{"Small" characteristic length $r_{0}$ and the non-ideality
parameters $\gamma_{s,k}$} \label{sec:rd1}

{\bf The connection of $r_{0}$ with Landau radius $r_{L}$.} As it
was announced in Part~1, all characteristic lengths of the developed
method appear already in the case of single-component systems. The
first of them is "small" characteristic length $r_{0}$, given by
Eqs.~(\ref{eq261})-(\ref{eq271}), which is interpreted as a radius
of the sphere centered on the probe particle and classically
forbidden for other free particles. Because of that it was useful to
compare the radius $r_{0}$ with a known characteristic length which
has similar physical sense. Here, we keep in mind Landau's radius
\begin{equation}
\label{eq33a} r_{L}  = \frac{(Ze)^2}{kT}
\end{equation}
which is used in the case of the classical systems (see e.g.
\cite{ebe76}). In accordance with Eqs.~(\ref{eq261})-(\ref{eq271})
we have that
\begin{equation}
\label{eq33}  r_{0}  =
\frac{1}{3}\kappa^{2}r_{s}^{3}\cdot\left(1+O(x^{3})\right), \qquad x
\ll 1,
\end{equation}
where in the classical case ($\partial \mu/\partial N = kT/N$) the
relation
\begin{equation}
\label{eq33a} \frac{1}{3}\kappa^{2}r_{s}^{3}=\frac{(Ze)^2}{kT}
\end{equation}
is valid. So, in this case the radii $r_{0}$ and $r_{L}$ are
connected by relation
\begin{equation}
\label{eq34a}  \lim_{x \to 0}\frac{r_{0}}{r_L} = 1.
\end{equation}
It means that $r_L$ represents an approximation od the
characteristic length $r_{0}$ which is applicable in the region of
small $x$. Consequently, {\it the parameter $r_{0}$ can be treated
as the generalization of Landau's characteristic length} $r_{L}$
{\it which introduces into consideration from physical reasons}. Let
us emphasize that, contrary to $r_{L}$ which is principally
unlimited, the radius $r_{0} < r_{s}$ for any $\kappa r_{s} > 0$, in
accordance with the conditions in real physical systems require, and
that Wigner-Seitz's radius $r_{s}=\lim_{x \to \infty}r_{0}$.

{\bf The connection of $\gamma_{s,k}$ with non-ideality parameters
$\Gamma$ and $\gamma$.} From Eqs.~(\ref{eq261})-(\ref{eq271}),
(\ref{eq33}) and (\ref{eq34a}) follow the expressions
\begin{equation}
\label{eq36}  \gamma _{s} = \frac{r_{0}}{r_{s}}, \qquad \gamma
_{\kappa } = \frac{r_{0}}{r_{\kappa}},
\end{equation}
for the coefficients $\gamma_{s}$ and $\gamma_{k}$. In connection
with these expressions in the classical case one should keep in mind
the relation (\ref{eq34a}) and the fact that $r_{\kappa}=r_{D}$,
where $r_{D}$ is the corresponding Debye's radius. With respect to
this it is useful to compare Eqs.~(\ref{eq36}) with the expressions
for known quantities
\begin{equation}
\label{eq37}  \Gamma = \frac{r_{L}}{r_{s}}=\frac{\left( {Ze}
\right)^2}{kT \cdot r_s }, \quad \gamma =
\frac{r_{L}}{r_{D}}=\frac{\left( {Ze} \right)^2}{kT \cdot r_{D} },
\end{equation}
which usually use as the non-ideality parameters for the classical
systems. Usually the parameters $\Gamma$ and $\gamma$, similarly to
Landau's radius $r_{L}$, introduce from some physical reasons (see
for example \cite{ebe76,kra86}). From
Eqs.~(\ref{eq261})-(\ref{eq271}), (\ref{eq36}) and (\ref{eq37}) it
follows that in the classical case
\begin{equation}\label{eq38a}
\lim_{x \to 0}\frac{\gamma_{s}}{\Gamma} = 1, \qquad \lim_{x \to
0}\frac{\gamma_{\kappa}}{\gamma} = 1,
\end{equation}
where $x=\kappa r_{s}$. It means that the parameters $\Gamma$ and
$\gamma$ appear here as approximative values of the coefficients
$\gamma _{s} $ and $\gamma _{\kappa } $ in the classical case in the
region $x \ll 1$. Consequently, {\it the parameters} $\gamma _{s} $
{\it and} $\gamma _{\kappa } $ {\it can be treated as the
generalized non-ideality parameters of single-component systems with
any} $N$, $T$ {\it and} $Z$ {\it which make possible their
non-relativistic treatment}.

From Eqs.~(\ref{eq251}) and (\ref{eq271}) it follows that the
coefficients $\gamma _{s} $ and $\gamma _{\kappa }$ represent the
functions of the parameter $r_{s}/r_{\kappa}$ which is closely
connected with another known classical quantity. Namely, in the
classical case ($r_{\kappa}=r_{D}$) we have that
$(r_{s}/r_{D})^{-3}=n_{D}$, where $n_{D}$ is so called Debye's
number, i.e. the mean number of the free particles in the sphere
with the radius $r_{D}$.

\begin{figure}[htbp]
\centerline{\includegraphics[width=\columnwidth,
height=0.75\columnwidth]{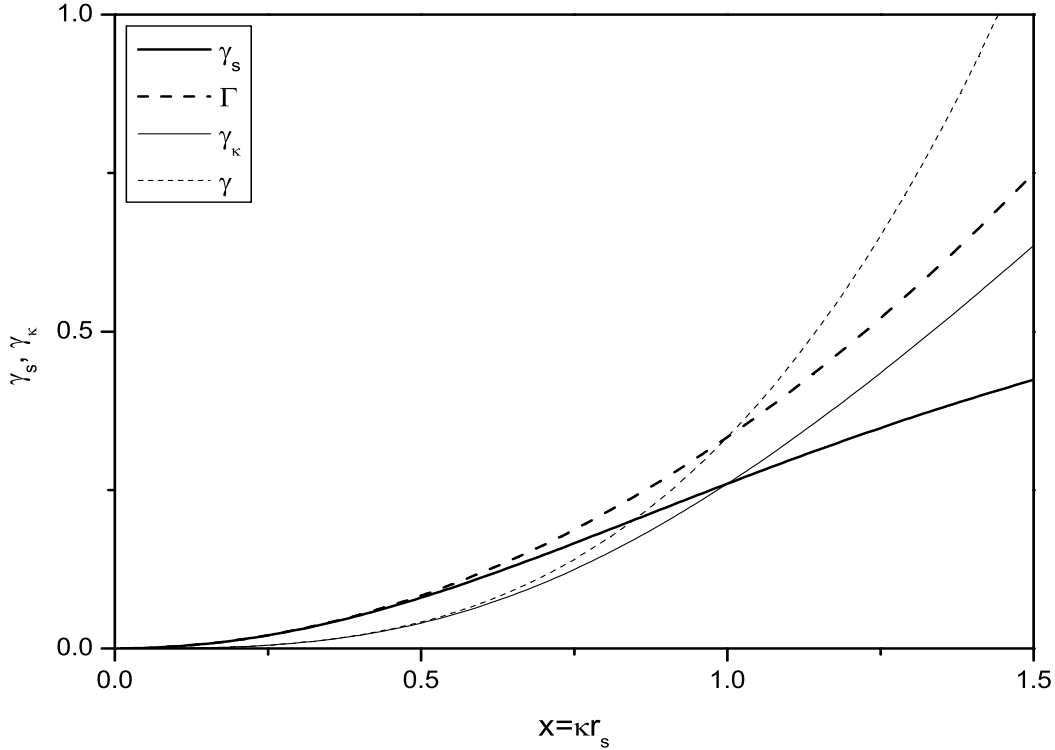}} \caption{The behavior of the
parameters $\gamma _{s} = r_0 / r_s $ and $\Gamma=r_{L}/r_{s}$, and
$\gamma _{\kappa } = r_0 / r_\kappa $ and $\gamma=r_{L}/r_{D}$,
defined by Eqs.~(\ref{eq261})-(\ref{eq271}) and (\ref{eq37}), as
functions of $x$, in the classical case.} \label{fig:gamma}
\end{figure}

The behavior of the coefficients $\gamma _{s} $ and $\gamma _{\kappa
}$ and the parameters $\Gamma$ and $\gamma$ as functions of $\kappa
r_s $ is illustrated in Fig.~\ref{fig:gamma}. This figure shows that
in the region $\kappa r_{s} \le 0.5$ the values of parameters
$\Gamma$ and $\gamma$ are practically same as the values of $\gamma
_{s} $ and $\gamma _{\kappa }$ and. They values left very closed up
to the value $\kappa r_{s} =1$.

\subsection{"Medium" characteristic length $r_{c}$ and the radius $r_{\kappa}$}
\label{sec:gg1}

{\bf The charges $Q_{in,out}(r)$.} The electro-neutrality condition
Eq.~(\ref{eq42}) can be presented in the form:
$Q_{in}(r)+Q_{out}(r)=0$, where  $0 \le r < \infty$, and the
quantities $Q_{in,out}(r)$ are given by expressions
\begin{equation}
Q_{in} (r) = Ze + \int\limits_0^r {\rho (r')4\pi r'^2dr'}, \qquad
Q_{out} (r) = \int\limits_r^\infty {\rho (r')4\pi r'^2dr'},
\label{eq16}
\end{equation}
with the charge density $\rho(r)$ defined by Eq.~(\ref{eq21}). One
can see that $Q_{in}(r)$ and $Q_{out}(r)$ represent the total charge
inside the sphere with radius $r$, centered in the point $O$, and
the total charge of the rest of the space, respectively.

Probably, the quantities $Q_{in,out}(r)$ could have different
applications. So, the fact that in the considered case the total
average charge of the probe particle self-sphere, i.e.
$Q_{in}(r=r_{s})$, is strictly equal to the average charge of the
free particles inside this sphere, can be of interest. However, in
the further consideration the quantity $Q_{in}(r)$ will play only an
accessory role. Namely, it will be used only the fact that
$Q_{in}(r)$, obtained by means Eqs.~(\ref{eq21}) and (\ref{eq16}),
has the physical sense in the whole space (including the region $r <
r_{s}$), and can characterize the degree of the probe particle
charge screening.

{\bf The connection of $r_{c}$ with the radius $r_{\kappa}$.} In the
case of single-component systems the screening constant $\kappa$,
defined by Eq.~(\ref{eq181}), represents one of the main parameters
in both methods - the method developed in this work and
Debye-H\"{u}ckel's (DH) method. Consequently, the characteristic
lengths $r_{\kappa}=1/\kappa$ has to be treated in the similar way.

Within DH method it is usual to interpret $r_{\kappa}$, which is
equal to Debye's radius $r_{D}$ in the classical case, as the
screening radius. That is based on the fact that $r_{\kappa}$ is a
distance from the probe particle where the DH electrostatic
potential (described in Part~1) becomes less than Coulomb potential
of this particle by the factor $e^{ - 1}$.

\begin{figure}[htbp]
\centerline{\includegraphics[width=\columnwidth,
height=0.75\columnwidth]{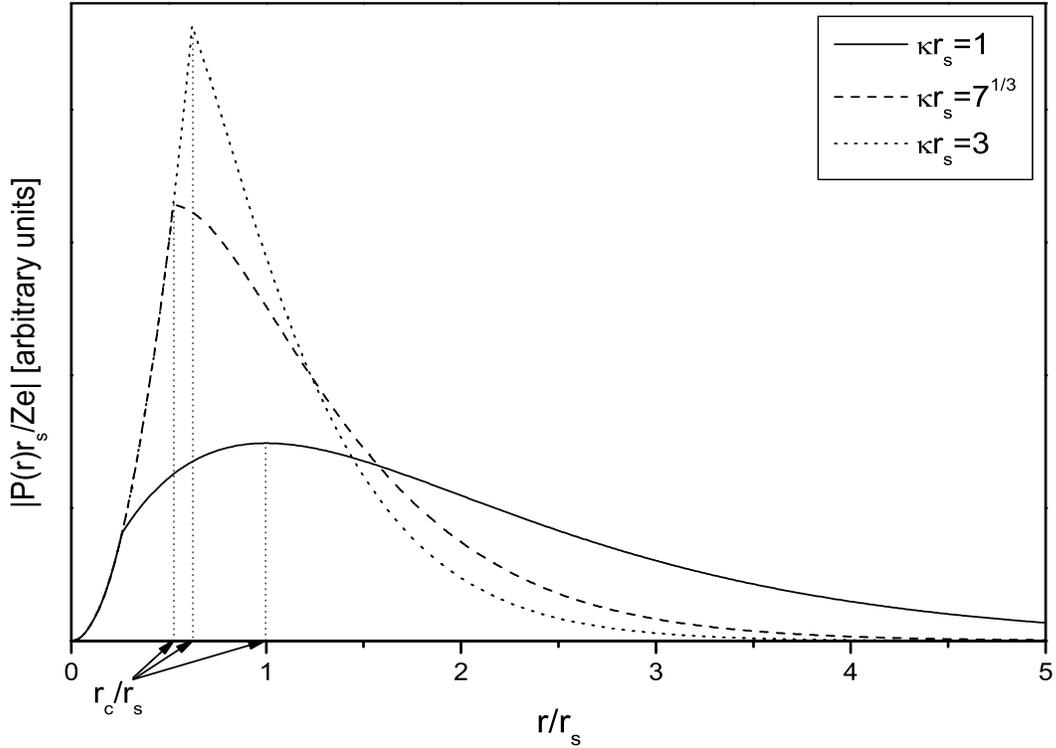}} \caption{The behavior of the
radial charge density $P(r)=4\pi r^{2} \rho(r)$ for the cases:
$\kappa r_s = 1$, when $\kappa r_0 < 1$; $\kappa r_s = 7^{{1 \over
3}}$, when $\kappa r_0 = 1$; $\kappa r_s = 3$, when $\kappa r_0 >
1$. The points where $P(r)$ has the maximal value are denoted by
$r_{c}$.} \label{fig:P}
\end{figure}

Here we will consider the role of radius $r_{\kappa}$ from other
aspect in order to clarify its real physical sense in both methods
and determine the region of its applicability. For this purpose, we
will introduce the radial charge density $P(r)=4\pi r^2 \cdot \rho
(r)$, where $\rho (r)$ is given by Eq.~(\ref{eq21}), and the
characteristic length $r_{c}$ defined by relation
\begin{equation}
\label{eq38}  |P(r_{c})|=\max\limits_{0 < r < \infty}|P(r)|.
\end{equation}
The behavior of $P(r)$ for several values of $x=\kappa r_{s}$ is
shown in Fig.~\ref{fig:P}. This figure illustrates the fact that for
$\kappa r_{s} < 7^{1 \over 3}$ the point $r=r_{c}$ there is in the
region $r > r_{0}$, while for $\kappa r_{s} \ge 7^{1 \over 3}$ we
have that $r_{c}=r_{0}$. Since in the region $r > r_{0}$ the charge
density $\rho(r) \sim \exp(-\kappa r)/r$, the parameter $r_{c}$ for
$\kappa r_{s} < 7^{1 \over 3}$ represents the root of equation
\begin{equation}
\label{eq39}  \frac{dP(r)}{dr} = 0,
\end{equation}
where $P(r)\sim r \cdot \exp(-\kappa r)$. Consequently, in this
region the parameter $r_{c}$ is equal to $r_{\kappa}$. It means that
within the method developed in this work we have the relations
\begin{equation}
\label{eq241}  r_{c} = \left\{ {{\begin{array}{*{20}c}
 {r_\kappa ,} \hfill & {0 < \kappa r_{s} < 7^{1 \over 3}, } \hfill \\
 {r_0 ,} \hfill & {7^{1 \over 3} \le \kappa r_{s} < \infty, } \hfill \\
\end{array} }} \right.
\end{equation}
which determine the real physical sense of the radius $r_{\kappa}$.
Namely, from Eq.~(\ref{eq241}) it follows that $r_{\kappa}$ has the
physical sense only in the region $\kappa r_{s} < 7^{1 \over 3}$
where $r_{\kappa}=r_{c}$, while in the region $\kappa r_{s} > 7^{1
\over 3}$ the length $r_{\kappa}$ loses any connection with the
charge distribution in the probe particle neighborhood and,
consequently, loses any physical sense. {\it Let us emphasize that
on the base of above mentioned one can conclude that it is always
$r_{c} \ge r_{0}$}.

Then, we will consider the problem of screening of the probe
particle as a problem of compensation of its charge $Ze$ within the
sphere of a radius $r$ centered at the probe particle. For that
purpose we will draw attention to the fact that uncompensated part
of this charge in the case of such a sphere is equal to the charge
$Q_{in} (r)$. Keeping this in mind, we compared the charge $Q_{in}
(r_{\kappa})$ with the charge $Ze$ in the region $r_k > r_0 $. Using
the relations (\ref{eq16}) and (\ref{eq21}) it could be shown that
in this whole region $Q_{in} (r_{\kappa}) > 0.735\cdot Ze$. From
here it follows that {\it the radius} $r_{\kappa}$, {\it as well as
the radius} $r_{c}$, {\it cannot be treated as a characteristic
length of full screening (neutrality) of the probe particle charge}.
Because of that, such a characteristic length is determined here in
another way.

\subsection{"Large" characteristic length $r_{n}$}
\label{sec:ncl}

{\bf The charge $Q_{in}(r)$ and the quantity $\nu (r)$.} In
accordance with Eq.~(\ref{eq21}) and the condition Eq.~(\ref{eq41}),
the charge $Q_{in}(r)$ can be presented in the form
\begin{equation}
\label{eqnun}  Q_{in}(r) = -Q_{out}(r) = Ze \cdot \nu(r),
\end{equation}
where the quantity $\nu(r)$ is given by the expression
\begin{equation}
\label{eqnun1} \nu(r)=\int\limits_{r}^{\infty} {\left[ {N - n(r')}
\right] \cdot 4\pi r'^{2}} dr',
\end{equation}
and $n(r)$ is the free particle density in the considered system
with the probe particle.

Although the expressions (\ref{eqnun1}) for $\nu(r)$ can be
determined for any $r > 0$, this quantity has a special physical
meaning in the region $r \ge r_{s}$. Namely, it can be shown that in
this region $\nu(r)$ represents the mean number of particles whose
surplus into the mentioned sphere of radius $r$ and corresponding
deficit in the rest of the space causes the deviations of
$Q_{in}(r)$ and $Q_{out}(r)$ from zero. Because of that only the
quantities $Q_{in}(r)$ and $\nu(r)$ in the region $r_{s} \le r \le
\infty$ will be needed in this paper. From this reason in further
considerations we will use the expression for $\nu(r)$ which is
applicable in the region $r_{0} < r < \infty$, since it is always
$r_{0} < r_{s}$. In accordance with Eq.~(20) for $n(r)$ from Part~1
we have that
\begin{equation}
\label{eq24App}  \nu(r) = \displaystyle{ {\left( {1 + \kappa r}
\right)\exp \left( { - \kappa r} \right) \cdot \chi (x),}} \qquad {r
> r_0 ,}
\end{equation}
where $x=\kappa r_{s}$, and $\chi(x)$ is defined by Eq~(30) and
illustrated by Fig.~3 from Part~1.

Similarly to $Q_{in}(r)$, the quantity $\nu(r)$ could have also some
different applications. For example, knowing of $\nu(r=r_{s})$ makes
possible to estimate the density $N_{p}$ of pairs of particles with
the inter-particle distance less then $r_{s}$. Namely, in the binary
approximation (not more than one particle inside the self-sphere of
any particle in the initial system) it can be shown that:
$N_{p}\cong (1/2)\cdot N \cdot \nu(r_{s})$, where $\nu(r_{s})$ is
given by Eq.~(\ref{eq24App}) with $r=r_{s}$. However, in further
consideration the charge $Q_{in}(r)$ and the quantity $\nu(r)$ will
play only an accessory role.

{\bf The characteristic length $r_{n}$ as the neutrality radius.} In
order to determine the required "large" characteristic length one
should consider the charge $Q_{in} (r) =  Ze \cdot \nu (r)$ in the
region $r \ge r_{s}$, where $\nu(r)$ is given by (\ref{eq24App}) and
has the sense which is described above. Let us remind that in this
region the charge $Ze$ of the probe particle is already completely
compensated by background charge of the probe particle self-sphere.
Consequently, in the region $r \ge r_{s}$ the charge $Q_{in}(r)$
represents a quantitative characteristic of deviation of the
neutrality of the sphere with radius $r$, centered at the prove
particle, which is exceptionally caused by the charge of $\nu(r)$
particles which are enter in this sphere from the rest of space.

\begin{center}
\begin{figure}[htbp]
\centerline{\includegraphics[width=\columnwidth,
height=0.75\columnwidth]{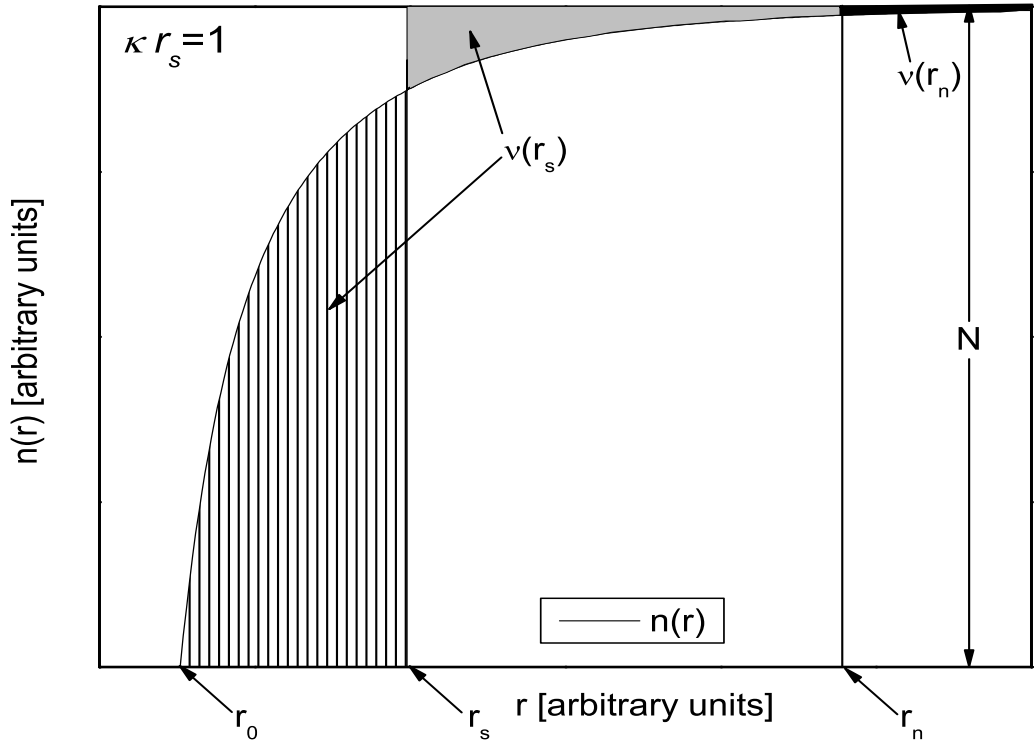}} \caption{The visualization of
the quantity $\nu (r) = Q_{in}(r)/Ze$ for $r = r_{s} $ and $r =
r_{n} $.} \label{fig:nu}
\end{figure}
\end{center}

From (\ref{eqnun}) and (\ref{eq24App}) it could be seen that
$|Q_{in}(r)|$ in the region $r \ge r_s$ almost exponentially
decreases from the maximum value $|Q_{in}(r_{s})|$ down to zero,
with the increasing of $r$. Consequently, as the requested screening
length, denoted here as $r_{n} $, can be taken the root of equation
\begin{equation}
\label{eq44}  \frac{\vert Q_{in} ( r )\vert}{\vert Q_{in} (r_{s} )
\vert} = e^{ - 1},
\end{equation}
where $Q_{in}(r)$ is given by (\ref{eqnun}) and (\ref{eq24App}). In
accordance with this $r_{n}$ can be treated as the neutrality
radius. Here, $r_{n}$ is the radius of such a sphere centered at the
probe particle for which charge exchange with the rest of the space
becomes practically negligible in comparison with the similar
exchange in the case of probe particle self-sphere, which is
illustrated by Fig.~\ref{fig:nu}. In the case of the initial system
(see Part~1) the radius $r_{n}$ can be interpreted as a radius of
minimal sphere which can be considered as practically neutral one.

\begin{figure}[htbp]
\centerline{\includegraphics[width=\columnwidth,
height=0.75\columnwidth]{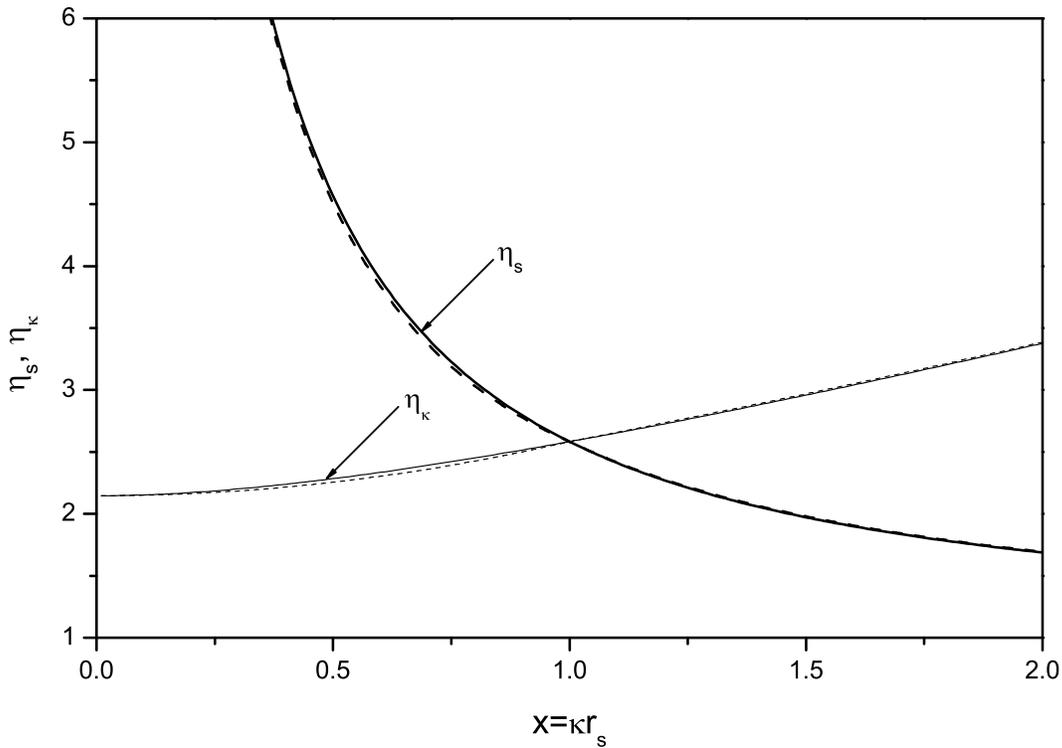}} \caption{The behavior of the
coefficients $\eta _{s} = r_{n}/r_s$, $\eta _{\kappa } =
r_{n}/r_\kappa$. Full lines show $\eta _{s}$ and $\eta _{\kappa}$
numerically determined from equation (\ref{eq45}); dashed lines -
the same quantities determined by approximate expressions
(\ref{eq47}) and (\ref{eq48}).} \label{fig:eta}
\end{figure}

From (\ref{eq44}) and (\ref{eq24App}) it follows that the
characteristic length $r_{n} $ represents the root of the equation:
\begin{equation}
\label{eq45}  \frac{1 + \kappa r} {1 + \kappa r_{s}} \cdot \exp [ -
\kappa ( r - r_{s} )] = e^{ - 1},
\end{equation}
which can be determined only numerically. Here, it is presented in
two equivalent form, namely
\begin{equation}
\label{eq46}  r_{n} = r_s \cdot \eta _{s} ,\qquad r_{n} = r_\kappa
\cdot \eta _{\kappa } ,
\end{equation}
where the coefficients $\eta _{s} $ and $\eta _{\kappa } $ are taken
in the form
\begin{equation}
\label{eq46a}  \eta _{s} = [1 + x + g(x)]/x, \qquad \eta_{\kappa } =
1 + x + g(x),
\end{equation}
since the member $g(x)$ can be very well approximated by means of
the easy expressions
\begin{equation}
\label{eq46b}  g(x) = \left\{ {{\begin{array}{*{20}c} \displaystyle{
1.14619 - x + 0.43688 \cdot x^2 } \hfill & {0 \le x \le
1,} \hfill \\
\displaystyle{1/x - 0.5/x^{2} + 0.08307/x^3}
\hfill & {1 \le x < \infty ,} \hfill \\
\end{array} }} \right.
\end{equation}
where at the point $x = 1$ both expressions give the same value
$g(1) = 0.58307$. For the coefficients $\eta _{s} $ and $\eta
_{\kappa } $ we obtain then following approximate expressions
\begin{equation}
\label{eq47}  \eta _{s} = \left\{ {{\begin{array}{*{20}c}
\displaystyle{ {\frac{2.14619}{x} + 0.43688 \cdot x,}} \hfill & {0 <
x
\le 1,} \hfill \\
\displaystyle{ {1 + \frac{1}{x}\left( {1 + \frac{1}{x} -
\frac{0.5}{x^2} +
\frac{0.08307}{x^3}} \right),}} \hfill & {1 \le x < \infty ,} \hfill \\
\end{array} }} \right.
\end{equation}
\begin{equation}
\label{eq48}  \eta _{\kappa } = \left\{ {{\begin{array}{*{20}c}
 {2.14619 + 0.43688 \cdot x^2,} \hfill & {0 < x \le 1,} \hfill \\
\displaystyle{ {x + 1 + \frac{1}{x}\left( {1 - \frac{0.5}{x} +
\frac{0.08307}{x^2}}
\right),}} \hfill & {1 \le x < \infty ,} \hfill \\
\end{array} }} \right.
\end{equation}
where $x=\kappa r_{s}$. The behavior of $\eta _{s} $ and $\eta
_{\kappa } $ obtained numerically from basic equation (\ref{eq45})
and by approximative expressions (\ref{eq46a}) and  (\ref{eq46b}),
is illustrated in Fig.~\ref{fig:eta}. This figure shows that the
mentioned approximative expressions give very good results. From
(\ref{eq47}) it follows that: $\mathop {\lim }\limits_{x \to 0} \eta
_{s} = \infty $ and $\mathop {\lim }\limits_{x \to \infty } \eta
_{s} = 1$. Consequently, on the base of (\ref{eq46}) we have it
that: $r_s < r_{n} < \infty $ for any $\kappa r_s > 0$, and
Wigner-Seitz's radius $r_{s}=\lim_{x \to \infty}r_{n}$ . Then, from
(\ref{eq48}) it follows that: $\eta _{\kappa }
> 1$ for $0 < \kappa r_s < \infty $. {\it On the base of (\ref{eq46}) we
have it that: $r_{n} > r_{c}$ for any $\kappa r_s > 0$}.

\section{The screening parameters of the two-component systems} \label{sec:nip2}

\subsection{"Small" characteristic lengths $r_{0;i,e}$ and non-ideality parameters
$\gamma_{s;i,e}$ and $\gamma_{\kappa;i,e}$}

{\bf The connection of $r_{0;i,e}$ with Landau radii $r_{L;i,e}$}.
In Part~2 it was shown that the simplifications which give the
possibility to describe a two-component plasma in DH method by means
of unique screening constant are unacceptable and that its ion and
electron components have to be described by own screening constant
$\kappa_{i}$ and $\kappa_{e}$ given by Eqs.~(\ref{eq292}) and
(\ref{eqalf}). One can see that these screening constants are
substantially different from the DH screening constant (see Part~2).
In the general case $\kappa_{i} \neq \kappa_{e}$. Unique exception
is the case of completely classical plasma with $Z_{i}=1$, and
$T_{i}=T_{e}$, but even in this case the common value of ion and
electron screening constants is significantly different from DH
screening constant.

The existence of two special screening constants $\kappa_{i}$ and
$\kappa_{e}$ causes that in the case of two-component plasma we have
two groups of screening parameters, analogous to the screening
parameters described in the previous Section, but for ion and
electron components separately (see Part~2). As first, we will
consider "small" characteristic lengths $r_{0;i}$ and $r_{0;e}$,
which are analogous to the characteristic length $r_{0}$ in the
single-component case, and have the sense of radii of the spheres
centered on the probe particles which are classically forbidden for
the ions in the case (i) and for electrons in the case (e).

Keeping in mind Eqs.~(\ref{eq251}) and (\ref{eq702}), as well as the
facts that $(1- \alpha) \cong 1$ in the case of weakly non-ideal
plasma and that in the classical case $\partial \mu_{i,e}/\partial
N_{i,e} = kT_{i,e}/N_{i,e}$, we have the relation
\begin{equation}
\label{eq34b}  \lim_{x_{i,e} \to 0}\frac{r_{0;i,e}}{r_{L;i,e}} = 1,
\end{equation}
where $x_{i,e}=\kappa_{i,e} r_{s;i,e}$, and $r_{L;i}$ and $r_{L;e}$
are known ion and electron Landau's radii, namely
\begin{equation}
\label{eq33b} r_{L;i}  = \frac{(Z_{i}e)^2}{kT_{i}}, \qquad r_{L;e} =
\frac{e^2}{kT_{e}}.
\end{equation}
One can see that $r_{L;i}$ and $r_{L;e}$ represent the approximation
of $r_{0;i}$ and $r_{0;e}$ in the classical case in the regions
$x_{i,e} \ll 1$. Consequently, in the case of two-component
classical plasma the characteristic lengths $r_{0;i}$ and $r_{0;e}$
can be treated as the corresponding generalization of Landau's radii
$r_{L;i}$ and $r_{L;e}$. It is important that, contrary to
$r_{L;i,e}$ which are principally unlimited, the radii $r_{0;i,e} <
r_{s;i,e}$ for any $\kappa_{i,e} r_{s;i,e} > 0$, and that
Wigner-Seitz's radii $r_{s;i,e}=\lim_{x_{i,e} \to \infty}
r_{0;i,e}$.

{\bf The connection of $\gamma_{s;i,e}$ and $\gamma_{\kappa;i,e}$
with non-ideality parameters $\Gamma_{i,e}$ and $\gamma_{i,e}$.} On
the base of Eqs.~(\ref{eq251}), (\ref{eq702}), (\ref{eq34b}) and
(\ref{eq33b}) in the case of classical plasma, we can obtain the
relations
\begin{equation}\label{eq38b}
\lim_{x_{i,e} \to 0}\frac{\gamma_{s;i,e}}{\Gamma_{i,e}} = 1, \qquad
\lim_{x_{i,e} \to 0}\frac{\gamma_{\kappa;i,e}}{\gamma_{i,e}} = 1,
\end{equation}
where $x_{i,e}=\kappa_{i,e} r_{s;i,e}$, and the quantities
$\Gamma_{i,e}$ and $\gamma_{i,e}$ are often used the classical ion
and electron non-ideality parameters, given by relations
\begin{equation}
\label{eq99a}  \Gamma_{i,e} = \frac{\left( {Z_{i,e}e}
\right)^2}{kT_{i,e} \cdot r_{s;i,e} }, \qquad \gamma_{i,e} =
\frac{\left( {Z_{i,e}e} \right)^2}{kT_{i,e} \cdot r_{\kappa;i,e} }.
\end{equation}
Let us emphasize that the parameters $\Gamma_{i,e}$ and
$\gamma_{i,e}$, similarly to $\Gamma$ and $\gamma$ in the
single-component case, are also introduced from "some physical
reasons" (see for example \cite{ebe76,kra86}. However, one can see
that $\Gamma_{i,e}$ and $\gamma_{i,e}$ represent the approximations
of the ion and electron non-ideality parameters $\gamma_{s;i,e}$ and
$\gamma_{\kappa;i,e}$ in the region $x_{i,e} \ll 1$. Consequently,
$\gamma_{s;i,e}$ and $\gamma_{\kappa;i,e}$ can be treated as the
generalization of ion and electron non-ideality parameters of
two-component classical plasma. The behavior of $\gamma_{s;i,e}$ and
$\gamma_{\kappa;i,e}$ as functions of $\kappa_{i,e} r_{s;i,e}$ is
similar to the behavior of analogous parameters in the case of
single-component system (see Fig.~\ref{fig:gamma}).

Let us draw attention that, {\it contrary to the single-component
case, the coefficients $\gamma_{s;i,e}$ and $\gamma_{\kappa;i,e}$
illustrate the physical inadequateness of Debye's parameters $r_{D}$
and $n_{D}$ for two-component systems.} Namely, in accordance with
Eqs.~(\ref{eq702}) and (\ref{eq712}) these coefficients represent
the functions of the parameters $r_{s;i,e}/r_{\kappa;i,e}$. However,
in two-component case we have that $r_{\kappa;i,e} \neq r_{D}$ and
the quantity $(r_{s;i,e}/r_{\kappa;i,e})^{-3} \neq n_{D}$.

\subsection{"Medium" characteristic lengths $r_{c;i,e}$} \label{sec:pc2}

{\bf The charges $Q_{in,out}^{(i,e)}(r)$.} Similarly to the
single-component case, we can take the electro-neutrality condition
(\ref{eq42}) in the form: $Q_{in}^{(i,e)}(r)+Q_{out}^{(i,e)}(r)=0$,
where  $0 \le r < \infty$, and the quantities $Q_{in}^{(i,e)}(r)$
and $Q_{out}^{(i,e)}(r)$ are given by relations
\begin{equation}
\label{eq18}  Q_{in}^{(i,e)} (r) = Z_{i,e} e + \int\limits_0^r {\rho
^{(i,e)}(r') \cdot 4\pi r'^2dr'} , \qquad Q_{out}^{(i,e)} (r) =
\int\limits_r^\infty {\rho ^{(i,e)}(r') \cdot 4\pi r'^2dr'},
\end{equation}
with the charge density $\rho ^{(i,e)}(r)$ defined by
Eq.~(\ref{eq22}). Similarly to the single-component case,
$Q_{in}^{(i,e)} (r)$ is the total charge of the whole sphere with
radius $r$, centered at the probe particle, and $Q_{out}^{(i,e)}(r)$
is the total charge of the rest of space.

In all further considerations it is needed to know the charges
$Q_{in}^{(i,e)}(r)$ only in the region $r_{s;i,e} \le r <infty$.
Keeping in mind Eq.~(\ref{eq21b}), as well as the fact that
$Q_{in}^{(i,e)}(r)=-Q_{out}^{(i,e)}(r)$ under the condition
(\ref{eq42}), we have that
\begin{equation}
\label{eq64}  Q_{in}^{(i,e)} (r) = Z_{i,e}e \cdot \left ( {1 -
\alpha } \right) \cdot \chi (x_{i,e} ) \left( {1 + \kappa _{i,e} r}
\right)\exp \left( { - \kappa _{i,e} r} \right),\quad r_{s;i,e} \le
r < \infty,
\end{equation}
where $x_{i,e}=\kappa_{i,e} r_{s;i,e}$, $Z_{e}=-1$ and $\chi (x)$ is
the same function as in Eq.~(\ref{eq24App}).

\begin{figure}[htbp]
\centerline{\includegraphics[width=\columnwidth,
height=0.75\columnwidth]{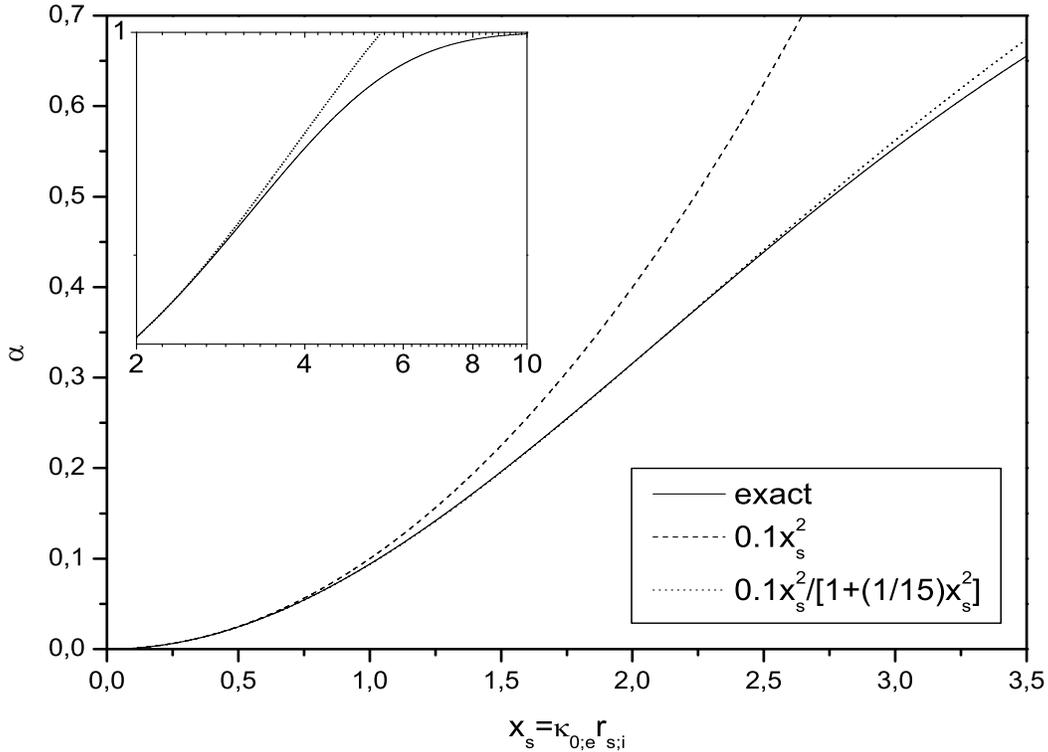}} \caption{The behavior of the
parameter $\alpha$, given by exact expression (\ref{eqalf}), and by
approximative expressions (\ref{eq50}) and (\ref{eq51}).}
\label{fig:alpha}
\end{figure}

For the practical applications of described method it is important
that the expression (\ref{eqalf}) for the coefficient of the
electron-ion correlation $\alpha$, which figure not only in
Eq.~(\ref{eq64}), but in expressions for all relevant quantities,
can be very well approximated by two simple expressions, namely
\begin{equation}
\label{eq50}  \alpha \cong \frac{1}{10} \cdot x_{s}^2, \qquad 0 <
x_{s} \lesssim \frac{3}{2},
\end{equation}
where the right side represents the first member of the expansion of
right side of Eq.~(\ref{eqalf}) in the series with the respect to
$x_{s}$, and
\begin{equation}
\label{eq51}  \alpha \cong \frac {1}{10}x_{s}^2 \cdot \left(1 +
\frac {1} {15}x_{s}^2 \right)^{-1}, \qquad 0 < x_{s} \lesssim
10^{1/2}.
\end{equation}
The behavior of the coefficient $\alpha$,  given by
Eqs.~(\ref{eqalf}), (\ref{eq50}) and (\ref{eq51}) is illustrated by
Fig.~\ref{fig:alpha}.

{\bf The connection of $r_{c;i,e}$ with the radii $r_{\kappa;i,e}$.}
In the two-component case the ion and electron screening constants
$\kappa_{i}$ and $\kappa_{e}$ have the sense of basic parameters of
the method developed in Part~2. Consequently, the corresponding
radii $r_{\kappa_{i}}=1/\kappa_{i}$ and
$r_{\kappa_{e}}=1/\kappa_{e}$ also represent the screening
parameters of this method. In order to clarify the real role of
$r_{\kappa_{i,e}}$ we will introduce the characteristic lengths
$r_{c;i}$ and $r_{c;e}$ which in the cases (i) and (e) have the
similar sense as the radius $r_{c}$ in the single-component case.

It can be shown that in the regions $0 < \kappa_{i,e} r_{s;i,e} < 1$
the parameters $r_{c;i,e}$ are the roots of equations
\begin{equation}
\label{eq39}  \frac{dP^{(i,e)}(r)}{dr} = 0,
\end{equation}
where $P^{(i,e)}(r) = 4\pi r^2 \cdot \rho^{(i,e)} (r)$, and
$\rho^{(i,e)} (r)$ is given by Eq.~(\ref{eq21b}). Consequently, we
have that in the region $0 < \kappa_{i,e} r_{s;i,e} \le 1$ the
relations
\begin{equation}
\label{eq39a}  r_{c;i,e}=r_{\kappa;i,e}, \qquad r_{c;i,e} >
r_{s;i,e}
\end{equation}
are valid. Due to the behavior of $\rho^{(i,e)}(r)$, which is
described in Part~2, we have it that in the case $\kappa_{i,e}
r_{s;i,e} \ge 1$ the parameter $r_{c;i,e} \le r_{s;i,e}$. In all
examined cases in the region $1 < \kappa_{i,e} r_{s;i,e} < \infty$
we obtain that
\begin{equation}
\label{eq39b}  r_{0;i,e} \le r_{c;i,e}, \qquad r_{c;i,e} \ne
r_{\kappa_{i,e}},
\end{equation}
except of the points $x_{i,e}=7^{1/3}$, where it is
$r_{c;i,e}=r_{\kappa;i,e}=r_{0;i,e}$. From just mentioned it follows
that the radii $r_{\kappa_{i}}$ and $r_{\kappa_{e}}$ have clear
physical sense only in the region $0 < \kappa_{i,e} r_{s;i,e} \le 1$
where it is defined by the relation (\ref{eq39a}).

Then, similarly to the single-component case we have to examine the
behavior of the charge $Q_{in}^{(i,e)}(r=r_{c;i,e})$ for
$\kappa_{i,e} r_{s;i,e} \le 1$. Since in this case $Q_{in}^{(i,e)}
(r)$ is given by Eq.~(\ref{eq64}), we have that
$Q_{in}^{(i,e)}(r_{c;i,e})
> 0.735 (1-\alpha) \chi(x_{i,e})\cdot Z_{i,e}e$. It means that
$Q_{in}^{(i,e)}(r_{c;i,e})$, in accordance with the behavior of
$\chi(x_{i,e})$, is comparable with the probe particle charge
$Z_{i,e}e$, excluding eventually the region $x_{i,e} \gg 1$.
Consequently, excluding the cases when $\alpha$ is close to unity,
the parameters $r_{c;i}$ and $r_{c;e}${\it cannot be treated as a
characteristic length of full screening (neutrality) of the probe
particle charges in the cases of (i) and (e)}. Because of that, such
characteristics lengths are determined here in the similar way as in
the single-component case.

Finally, we wish to draw attention to the fact that in the weakly
non-ideal plasma ($1- \alpha \cong 1 $) the radii $r_{\kappa;i}$ and
$r_{\kappa;e}$ are very closed to "medium" characteristic lengths in
the corresponding single-component systems (ion gas on the negative
background and electron gas on the positive background). This fact
has already been considered in connection with non-applicability of
DH method in the case of two-component plasmas. Namely, in
\cite{bow61,vit90,ada80,dju91,ada94}, where the case of the
classical plasma with $Z_{i}=1$ and $T_{i}=T_{e}$ were considered
(see Part~1), the screening constants and radius close to
$\kappa_{i,e}$ and $r_{\kappa;i,e}=1 / \kappa_{i,e}$ were used,
instead of DH screening constant $\kappa_{D}$ and radius
$r_{D}=1/\kappa_{D}$. The results obtained in this work justify such
a choice.

\subsection{"Large" characteristic lengths $r_{n;i}$ and $r_{n;e}$
as the neutrality radii}
\label{sec:nc2}

{\bf The expressions for $r_{n;i}$ and $r_{n;e}$.} Similarly to
Part~1, we will introduce the screening lengths $r_{n;i}$ and
$r_{n;e}$ which represent the roots of the equations
\begin{equation}
\label{eq100}  \frac{Q_{in}^{(i,e)}(r)}{Q_{in}^{(i,e)}(r_{s;i,e})} =
e^{ - 1},
\end{equation}
where the charge $Q_{in}^{(i,e)}(r)$ is given by Eq.~(\ref{eq64})
and one of Eqs.~(\ref{eqalf}), (\ref{eq50}) and (\ref{eq51}). From
it follows that Eq.~(\ref{eq100}) can be taken in the form
\begin{equation}
\label{eq102}  \frac{1 + \kappa_{i,e}r }{1 + \kappa_{i,e}r_{s;i,e}}
\cdot \exp \left[ { - \kappa_{i,e} (r - r_{s;i,e})} \right] = e^{ -
1},
\end{equation}
which is the same as the form of the corresponding equation for the
radius $r_{n}$ from previous Section. Consequently, we have that
$r_{n;i}$ and $r_{n;e}$ are given by relations
\begin{equation}
\label{eq104i}  r_{n;i} = r_{s;i} \cdot \eta _{s}(x_{i}) , \qquad
r_{n;e} = r_{s;e} \cdot \eta _{s}(x_{e}) ,
\end{equation}
\begin{equation}
\label{eq104e} r_{n;i} = r_{\kappa;i} \cdot \eta _{\kappa}(x_{i}),
\qquad r_{n;e} = r_{\kappa;e} \cdot \eta _{\kappa}(x_{e}),
\end{equation}
where the coefficients $\eta_{s}(x)$ and $\eta_{\kappa}(x)$ are
given by Eqs.~(\ref{eq46a})-(\ref{eq48}). On the base of these
expressions we have it that the relations $r_{n;i,e} > r_{s;i,e}$
and $r_{s;i,e}=\lim_{x_{i,e} to \infty} r_{n;i,e}$, as well as
$r_{n;i,e} > r_{c;i,e}$, are valid in the whole regions $0 <
\kappa_{i,e} r_{s;i,e} < \infty$.

\begin{figure}[htbp]
\centerline{\includegraphics[width=\columnwidth,
height=0.75\columnwidth]{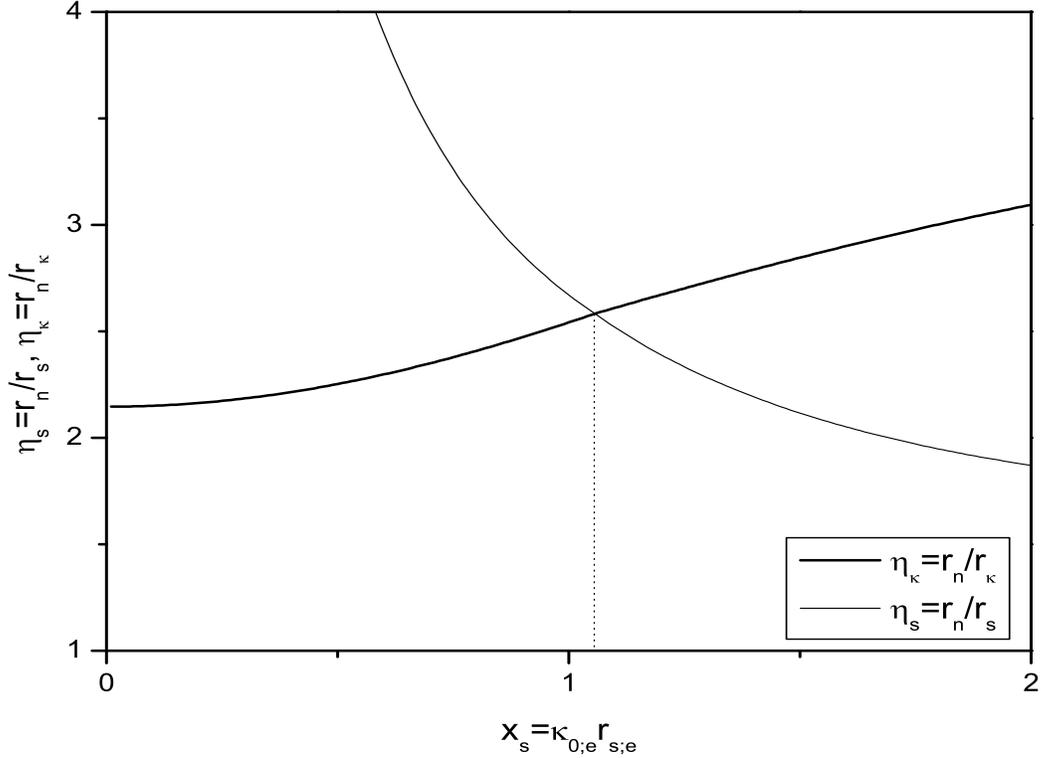}} \caption{The behavior of the
coefficients $\eta_{s}$ and $\eta_{\kappa}$ in the region $0 \le
x_{s} = \kappa_{0;e}r_{s;e} \le 2$ where the classical and general
expressions for these coefficients give practically the same
results. The case of plasma with $Z_{i}=1$ and $T_{i}=T_{e}$, when
$\kappa_{0;e}r_{s;e} = \kappa_{0;e}r_{s;i} \equiv x_{s}$ is
presented. The shift of the crossing point of the curves presented
related to the point $\kappa_{0;e}r_{s;e} = 1$ is caused by the
application of the general expressions for $\eta_{s}$ and
$\eta_{\kappa}$.} \label{fig:eta1}
\end{figure}

\begin{figure}[htbp]
\centerline{\includegraphics[width=\columnwidth,
height=0.75\columnwidth]{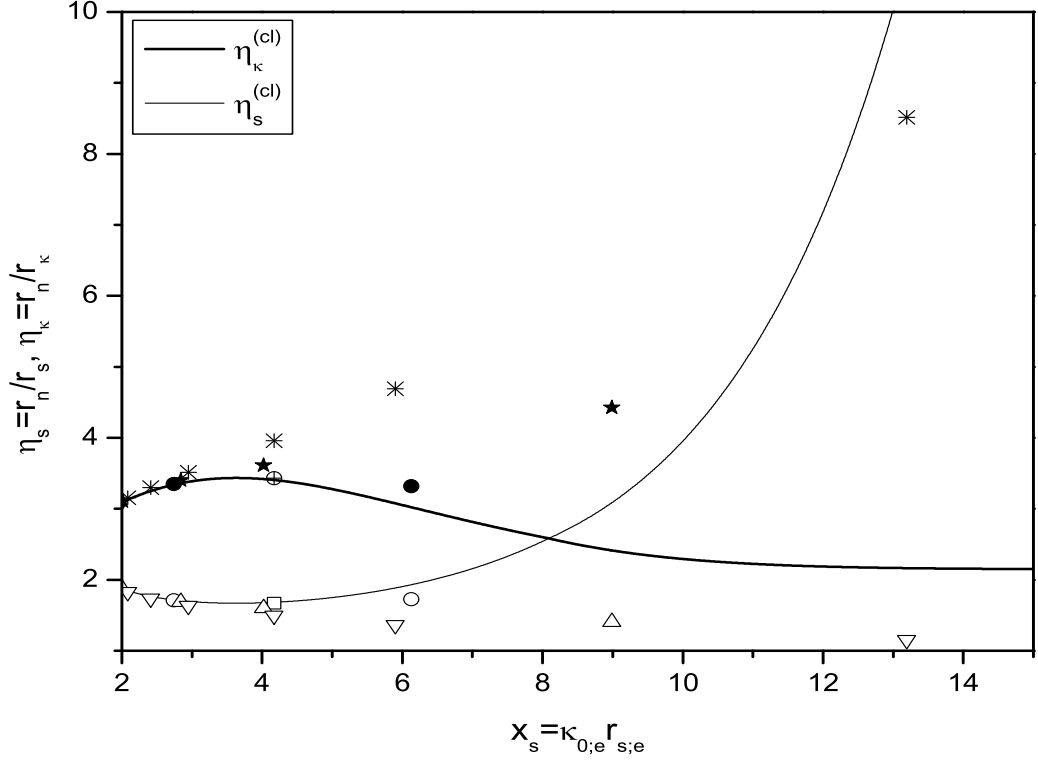}} \caption{The behavior of
coefficients $\eta_{s}$ and $\eta_{\kappa}$ in the region $x_{s} =
\kappa_{0;e}r_{s;e} > 2$. The curves $\eta_{s}^{(cl)}$ and
$\eta_{\kappa}^{(cl)}$ show the behavior of these coefficients
obtained in the classical case. The real behavior of the same
coefficients obtained with the general expression (\ref{eq42}) for
$\kappa_{0;e}$ is presented with: $\Box$- for $N_{e}=10^{19}
cm^{-3}$, $\circ$- for $N_{e}=10^{20} cm^{-3}$, $\bigtriangleup$-
for $N_{e}=10^{21} cm^{-3}$, and $\bigtriangledown$- for
$N_{e}=10^{22} cm^{-3}$ in the case $\eta_{s}$, and with: $\oplus$-
for $N_{e}=10^{19} cm^{-3}$, $\bullet$- for $N_{e}=10^{20} cm^{-3}$,
$\star$- for $N_{e}=10^{21} cm^{-3}$, and $\ast$- for $N_{e}=10^{22}
cm^{-3}$ in the case $\eta_{\kappa}$.} \label{fig:eta2}
\end{figure}

The behavior of the coefficients $\eta _{s}(x_{i,e})$ and $\eta
_{\kappa}(x_{i,e })$ in a wide region of $x_{i,e} = \kappa_{i,e}
r_{s;i,e} $, which are given by Eqs.~(\ref{eq292}), (\ref{eqalf}),
(\ref{eq712}), (\ref{eq46a})-(\ref{eq48}), (\ref{eq104i}) and
(\ref{eq104e}), is displayed in Figs.~\ref{fig:eta1} and
\ref{fig:eta2}. These figures clearly illustrate the changes of the
behavior of $\eta _{s}(x_{i,e})$ and $\eta _{\kappa}(x_{i,e })$ with
the increasing of deviation of the screening constant $\kappa_{i,e}$
from their classical values in the case $x_{i}=x_{e}$. Namely,
Fig.~\ref{fig:eta1} shows the behavior of $\eta _{s}(x_{i})$ and
$\eta _{\kappa}(x_{i})$ in the region $0 \le x_{s}=\kappa_{0,e}
r_{s;e} \le 2$ where the classical and general expressions give
practically the same results, while Fig.~\ref{fig:eta2} is related
to the region $2 < x_{s} \le 15$ where the classical and general
expressions give very different results.

\begin{figure}[htbp]
\centerline{\includegraphics[width=\columnwidth,
height=0.75\columnwidth]{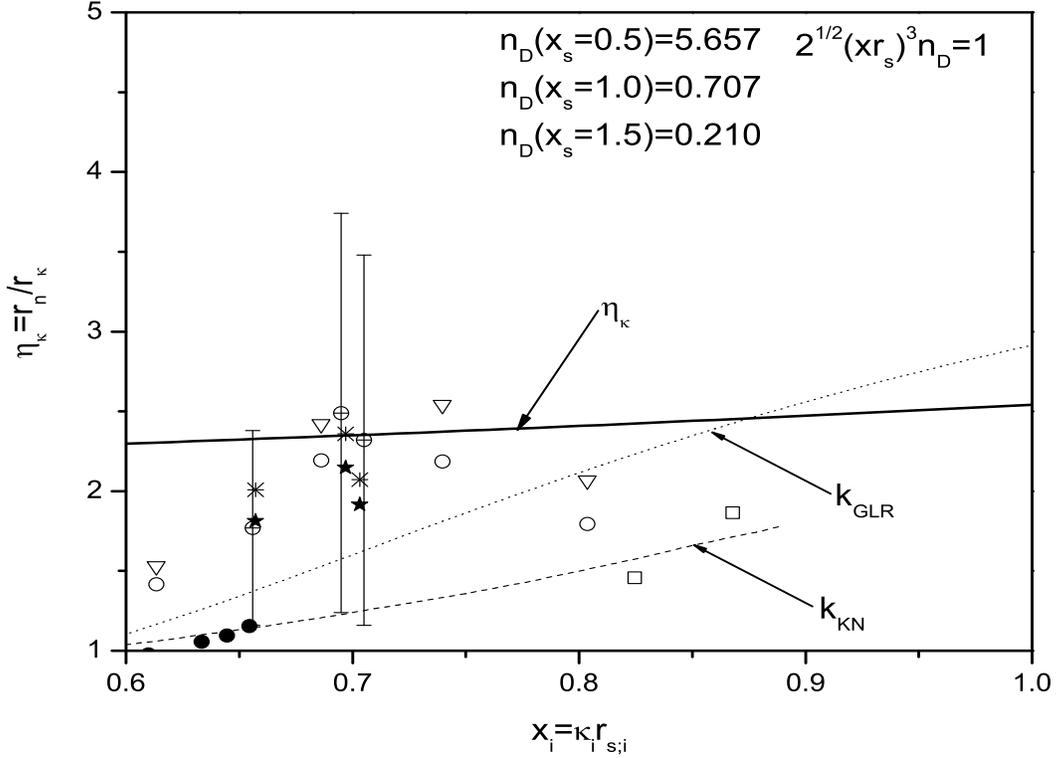}} \caption[a]{The comparison of
the coefficient $\eta_{\kappa}$ with the correction coefficients
$k_{c}=r_{scr}/r_{\kappa;i}$, where $r_{scr}$ is the effective
screening length determined in several papers on the base of
experimental data. The cases of completely classical plasmas with
$Z_{i}=1$ and $T_{i}=T_{e}$ are presented. The values of $k_{c}$ are
shown with: $\star$ and $\ast$- \cite{gun76,gun84},
$\bigtriangledown$ and $\circ$- \cite{gun83,gun84}, $\Box$-
\cite{gun72,gun84}, $\bullet$- \cite{gol78,gun84}. With $\oplus$ are
shown the values of $k_{c}$ for the same $N_{e}$ and $T_{e}$ as in
\cite{gun76}, but determined by means of expression for the plasma
conductivity from \cite{mih89}. The curves $k_{c;KN}$ and
$k_{c;GLR}$ show the behavior of $k_{c}$ determined by means of
analytical expressions from \cite{kak73} and \cite{gun83,gun84},
respectively.} \label{fig:etaexp}
\end{figure}

{\bf The applications and comparison with the existing experimental
data.} As we have already mentioned, the necessity of interpretation
of experimental data caused several attempts (see for example
\cite{kak73,gun84,vit01}) to determine of the characteristic
screening length $r_{scr}$ for two-component plasma which was taken
as: $r_{scr}=k_{c;D} \cdot r_{D}$, where $r_{D}$ is Debye's radius
for two-component system (see Part~2). In order to compare the
values of $r_{scr}$ and $r_{n}$ we take here $r_{scr}$ in the form
\begin{equation}\label{eqkc}
 r_{scr} = k_c \cdot r_{\kappa;i},
\end{equation}
where the correction factor $k_{c} = k_{c;D} r_{D}/r_{\kappa;i}$. We
will have in mind that the classical plasmas with $Z_{i}=1$ and
$T_{i}=T_{e}$ has been considered in the above mentioned papers, and
that in such plasmas: $r_{s;i}=r_{s;e}$, $\kappa_{i}=\kappa_{e}$,
$r_{\kappa;i}=r_{\kappa;e}$, $r_{n;i}=r_{n;e}$  and
$\eta_{\kappa;i}=\eta_{\kappa;e}$. The comparison of our results
with the existing experimental data is performed for several cases
and is presented in Fig.~\ref{fig:etaexp}. This figure shows the
behavior of the parameter $\eta_{\kappa;i}$ determined by
(\ref{eq104i})-(\ref{eq104e}) and the correction coefficients
$k_{c}$ taken from \cite{gun72,kak73,gun76,gol78,gun83,gun84} in the
region $0.6 \le \kappa r_{s} \le 1.0$. One can see a good agreement
with the correction coefficients obtained from several measurements
of conductivity from \cite{gun76,gun83}.

\begin{figure}[htbp]
\centerline{\includegraphics[width=\columnwidth,
height=0.75\columnwidth]{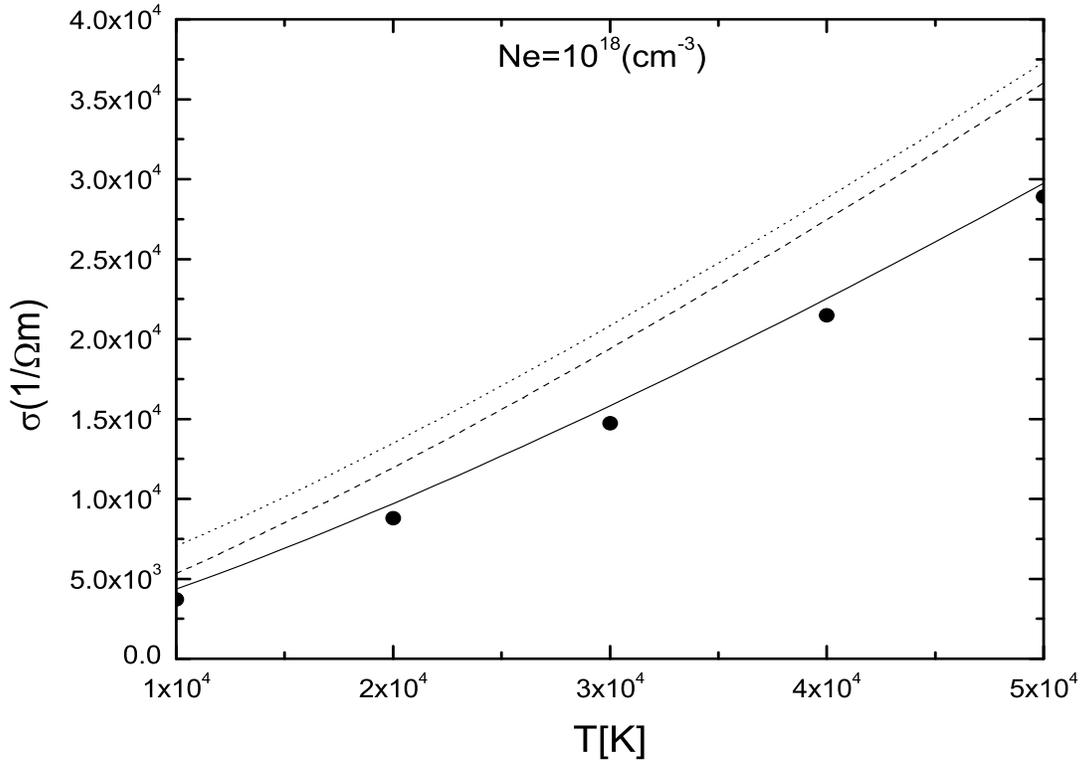}} \caption{The Spitzer's and
RPA conductivity of plasma with $Z_{i}=1$, $T_{i}=T_{e}=T$ and
$N_{e}=10^{18}(cm^{-3})$. The values of RPA conductivity ($\bullet$)
are taken form \cite{vit01}. Spitzer's conductivities are determined
by means of expression from \cite{gun76,gun84}, with the corrected
screening radius $r_{scr}$, and the corresponding calculations are
performed for: $r_{scr}=r_{D}$ - dotted curve; $r_{scr}=r_{D}\cdot
k_{KN}$ - dashed curve; $r_{scr}=r_{n;i}$ - full curve. Here $r_{D}$
is Debye's radius for two-component plasma (see Part~2), and
$k_{KN}$ is the corrected factor from \cite{kak73}.}
\label{fig:spe18}
\end{figure}

\begin{figure}[htbp]
\centerline{\includegraphics[width=\columnwidth,
height=0.75\columnwidth]{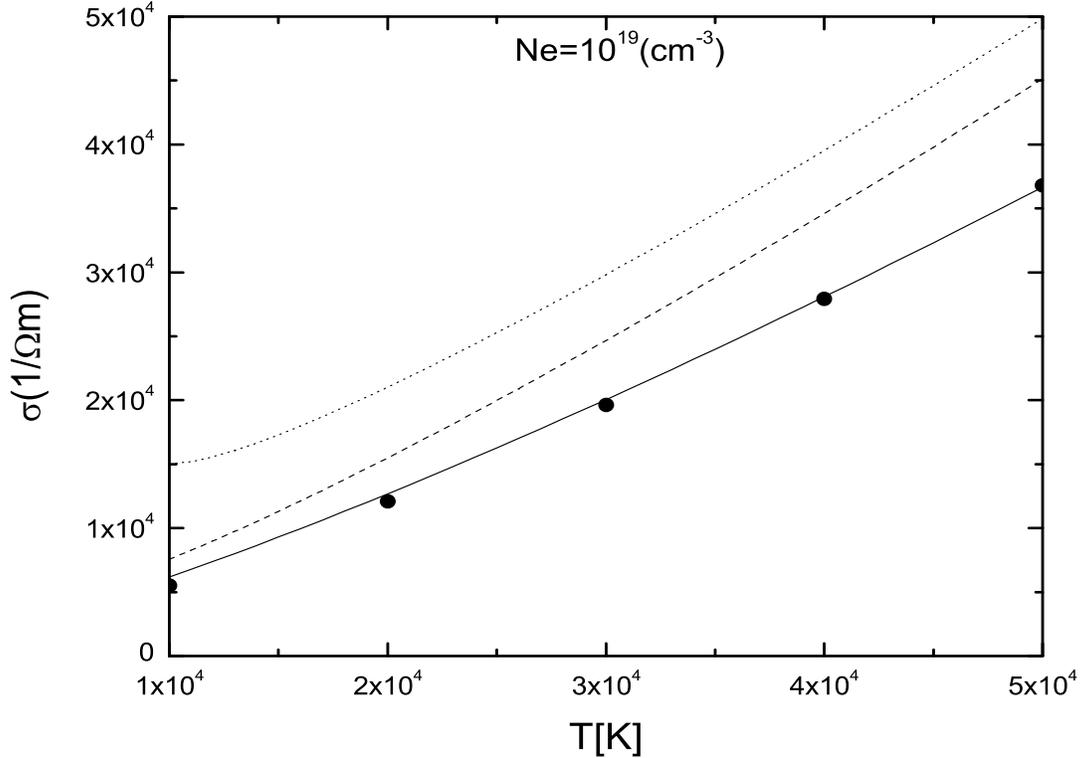}} \caption{Same as in
Fig.~\ref{fig:spe18}, but for $N_{e}=10^{19}(cm^{-3})$.}
\label{fig:spe19}
\end{figure}

In order to examine usage of $r_{n;i,e}$ as a screening radius in
the expressions of Spitzer's type for the conductivity of plasma, we
performed a calculation of conductivity for a fully ionized plasma
with $Z_{i}=1$ and $T_{i}=T_{e}=T$, where $r_{n;i}=r_{n;e}$. The
cases of plasmas with $N_{e}=10^{18}cm^{-3}$ and $10^{19}cm^{-3}$ in
the region $10^{4}K \le T \le 5\cdot 10^{4}K$ were considered. The
conductivities were determined by means of corrected Spitzer's
expression \cite{gun76} where Debye's radius $r_{D}$ for
two-component plasma is replaced (in so called Coulomb logarithm) by
screening radius $r_{scr}$. The calculation was performed for
$r_{scr}= r_{D}$, $r_{scr}=k_{KN} \cdot r_{D}$, where $k_{KN}$ is
the corrected factor from \cite{kak73}, and $r_{scr}=r_{n;i}$. The
results are presented in Figs.~\ref{fig:spe18} and \ref{fig:spe19}.
In the same figures the corresponding values of conductivity
determined by improved RPA method which is applicable for dense
non-ideal plasmas \cite{dju91,ada94,vit01} are also presented.

The Figs.~\ref{fig:spe18} and \ref{fig:spe19} show that Spitzer's
conductivity with $r_{scr}=r_{n;i}$ apparently converge to RPA
conductivity with the increasing of electron density $N_{e}$ from
$10^{18}cm^{-3}$ to $10^{19}cm^{-3}$, while Spitzer's conductivity
with $r_{scr}= r_{D}$ and $r_{scr}=k_{KN} \cdot r_{D}$ lie
significantly above. One can see that use of $r_{scr}=r_{n;i}$
provides the complete agreement with the results of RPA calculations
in the case $N_{e}=10^{19}cm^{-3}$. It is important that this
agreement becomes better when the non-ideality degree increases.

\section{Conclusions}
\label{sec:conc}

The new model method for describing of the electrostatic screening
in single- and two-components systems (electron-ion plasmas, dusty
plasmas, some electrolytes, etc.) developed in Part~1 and Part~2 of
this work, generates a group of new screening parameters. Here, we
keep in mind "small" characteristic length $r_{0}$ and the
non-ideality parameters $\gamma_{s}$ and $\gamma_{\kappa}$ in the
single-component case, and the corresponding characteristic lengths
$r_{0;i,e}$ and the non-ideality parameters $\gamma_{s;i,e}$ and
$\gamma_{\kappa;i,e}$ in the two-component case.

In connection with the mentioned screening parameters is established
that $r_{0}$ and $r_{0;i,e}$ represent the generalization of
classical Landau's radii $r_{L}$ and $r_{L;i,e}$, and
$\gamma_{s,\kappa}$, $\gamma_{s;i,e}$ and $\gamma_{\kappa;i,e}$ -
the generalization of known classical non-ideality parameters
$\Gamma$ and $\gamma$ in the single-component case, and
$\Gamma_{i,e}$ and $\gamma_{i,e}$ in the two-component case.

Apart of that, in this paper are introduced into consideration
"medium" and "large" characteristic lengths $r_{c}$ and $r_{n}$ in
the single-component case, and $r_{c;i,e}$ and $r_{n;i,e}$ in the
two-component case. The behavior of these radii is examined in the
whole regions $0 < \kappa r_{s} < \infty$ and $0 < \kappa_{i,e}
r_{s;i,e} < \infty$, where $\kappa$ and $\kappa_{i,e}$ are the
corresponding screening constants, and $r_{s}$ and $r_{s;i,e}$- the
corresponding Wigner-Seitz's radii. It was found that the considered
characteristic lengths satisfy the relations
\begin{equation}
r_{0} < r_{s} < r_{n}, \qquad r_{s} =\lim_{x \to \infty} r_{0}=
\lim_{x \to \infty} r_{n},
\end{equation}
\begin{equation}
r_{0;i,e} < r_{s;i,e} < r_{n;i,e}, \qquad r_{s;i,e} =\lim_{x_{i,e}
\to \infty} r_{0;i,e}= \lim_{x_{i,e} \to \infty} r_{n;i,e},
\end{equation}
\begin{equation}
r_{0} \le r_{c} < r_{n}, \qquad r_{0;i,e} \le r_{c;i,e} < r_{n;i,e},
\end{equation}
where $x=\kappa r_{s}$ and $x_{i,e}=\kappa_{i,e} r_{s;i,e}$. These
relations establish the two hierarchy systems of the characteristic
lengths, and causes a redefinition of Wigner-Seitz's radii as a
boundary screening lengths.

Then, it was found that the radius $r_{\kappa}=1/\kappa$ in the
single-component case has the sense only in the region of $x \le
7^{1/3}$, where $r_{\kappa}=r_{c}$, and the radii
$r_{\kappa;i,e}=1/\kappa_{i,e}$ have the sense only in the regions
$x_{i,e} \le 1$, where $r_{\kappa;i,e}=r_{c;i,e}$ in the
two-component case.

The results of application of the characteristic length $r_{n;i}$ as
the neutrality radius were compared in this paper with existing
experimental data in the cases of the classical plasmas with
$Z_{i}=1$ and $T_{i}=T_{e}$. It was found their  very good
agreement.

Finally, we wish to draw attention that developed method is suitable
for some astrophysical applications. Here we keep in mind that in
outer shells of stars the physical conditions change from those
which correspond to the rare, practically ideal plasma, to those
which correspond to extremely dense non-ideal one. However, the
method presented gives a possibility to describe the electrostatic
screening of all such outer shells in the same way, by means of the
obtained screening characteristics.

\begin{acknowledgments}
The authors are thankful to the University P. et M. Curie of Paris
(France) for financial support, as well as to the Ministry of
Science of the Republic of Serbia for support within the Project
141033 "Non-ideal laboratorial and ionospheric plasmas: properties
and applications".
\end{acknowledgments}

\begin{appendix}

\section{The lowering of the atomic ionization potential in plasma}
\label{sec:scon}

The method developed in Part~1 and Part~2 serve for describing of
the electrostatic screening in the single- and two-component systems
which contain only charged particles. However, as it is well known,
the presence of the neutral component in plasma can often be
neglected from the aspect of inner-plasma screening. It gives the
possibility to apply the results obtained in this work on the
problem of the lowering of the atomic ionization potential in
plasmas with the neutral component (see Part~1). In such a way we
will be able to compare the results obtained within the developed
and DH method. The results of this comparison will give another
important example of the inapplicability of DH method.

In this context we will remind that in Part~2 the potential energies
$U^{(i)}$ and $U^{(e)}$ of the ion and electron in plasma were
determined. In the case $Z_{i}=1$ and $T_{i}=T_{e}$, when
$U^{(i)}=U^{(e)}$, the ion potential energy $U^{(i)}$ was compared
with the corresponding DH ion potential energy $U_{D}^{(i)}$. Then,
we will denote with $\triangle I_{at}$ and $\triangle I_{at;D}$ the
atomic ionization potential determined within the developed and DH
method, and take into account that $\triangle I_{at}\sim U^{(i)}$
and $\triangle I_{at;D} \sim U_{D}^{(i)}$ where the proportionality
coefficients are equal or at least very closed. From here it follows
the relation
\begin{equation}\label{eqdelta}
 \frac{ \Delta I_{at}}{\Delta I_{at;D}} \cong
 \frac{U^{(i)}}{U_{D}^{(i)}}.
\end{equation}
The behavior of the right side of this relation is shown in Fig.~5
of Part~2. Accordingly to this figure we can conclude that $\Delta
I_{at}/\Delta I_{at;D} < 0.8$ in the whole region $0 <
\kappa_{i}r_{s;i} < \infty$. It is important to noted this fact,
since the handbooks (see also \cite{ebe76}), often used in
laboratories, recommend just DH lowering of the atom ionization
potential $\Delta I_{at;D}$, which would not be taken as one of
screening characteristics of electron-ion plasmas.

\end{appendix}

\newcommand{\noopsort}[1]{} \newcommand{\printfirst}[2]{#1}
  \newcommand{\singleletter}[1]{#1} \newcommand{\switchargs}[2]{#2#1}

\end{document}